\documentclass[runningheads]{llncs}
\usepackage{svg}
\usepackage{graphicx}
\usepackage{soul}
\usepackage{graphicx}
\usepackage{amsmath}
\usepackage{booktabs}
\usepackage{url}
\usepackage{amssymb}
\usepackage{amsmath}
\usepackage{wrapfig}
\usepackage{multirow}
\usepackage{graphicx}
\usepackage{subfigure}
\usepackage{makecell}
\usepackage{algorithm}  
\usepackage{algorithmic}  
\usepackage{booktabs}
\author{Submitted for Blind Review}
\begin{document}
	\title{Learning Recommender Systems with Soft Target: A Decoupled Perspective}
 \author{Hao Zhang\inst{1}, Mingyue Cheng\inst{1}\thanks{Corresponding author}, Qi Liu\inst{1}, Yucong Luo\inst{1}, Rui Li\inst{1}, Enhong Chen\textsuperscript{1}}
\institute{State Key Laboratory of Cognitive Intelligence, University of Science and Technology of China, Hefei, China \\ \email{\{zh2001, prime666, ruili2000\} @mail.ustc.edu.cn, \\ \{mycheng, qiliuql, cheneh\}@ustc.edu.cn} }
	%
	% \titlerunning{Abbreviated paper title}
	% If the paper title is too long for the running head, you can set
	% an abbreviated paper title here
	%
	\maketitle              % typeset the header of the contribution

		\begin{abstract}
		% Learning recommender systems with multi-class optimization objectives is a prevalent setting in the recommendation. However, vanilla softmax loss often ignores the difference between potential positive feedback and truly negative feedback. In this work, we propose a decoupled soft target optimization framework for recommender system, named DeSoRec, to enhance the learning of the user interests. To be specific, we first design a label propagation process to adaptively capture the potential preferences over unobserved items via the neighbors. Then, by carefully theoretical analysis of the traditional optimization framework using Kullback Leibler (KL) divergence and cross-entropy, we propose to decouple the objective into observed target and the remaining items to enhance the model learning. We conduct extensive experiments on four benchmark recommender models and four real world datasets, the experimental results demonstrate the effectiveness of the proposed approach. We will make our code publicly available after acceptance.
	Learning recommender systems with multi-class optimization objective is a prevalent setting in recommendation. However, as observed user feedback often accounts for a tiny fraction of the entire item pool, the standard Softmax loss tends to ignore the difference between potential positive feedback and truly negative feedback. To address this challenge, we propose a novel decoupled soft label optimization framework to consider the objectives as two aspects by  leveraging soft labels, including target confidence and the latent interest distribution of non-target items. 
 % Moreover, we find these two aspects are coupled in traditional soft-label optimization framework through our theoretical analysis, which limits the flexible adjustment of the weights of these two parts, potentially leading to sub-optimal performance. Hence, a more flexible loss function is designed to overcome this challenge. 
 Futhermore, based on our carefully theoretical analysis, we design a decoupled loss function to flexibly adjust the importance of these two aspects.
 To maximize the performance of the proposed method, we additionally present a sensible soft-label generation algorithm that models a label propagation algorithm to explore users' latent interests in unobserved feedback via neighbors. We conduct extensive experiments on various recommendation system models and public datasets, the results demonstrate the effectiveness and generality of the proposed method. \footnote{https://github.com/zh-ustc/desorec/}.
	\end{abstract}

	\section{Introduction}
	% Personalized recommendation 
	
	% Item recommendation with implicit feedback
	
	% In this work, we propose a novel optimization framework, named DeSoRec,
	%Personalized recommendation 
% Recommender system (RS)  \cite{liu2018illuminating,yu2018multiple} have become crucial tools for information filtering in various real world scenarios, including e-commerce, advertising, short video platform, and so on. As a data-driven technology, the key of the personalized RS  is to capture the users' preferences based on their historical interactions with items. 
% Among the user behaviors, implicit feedback such as clicks and purchases is typically leveraged to train the recommender systems ~\cite{wang2021denoising} due to its large volume. 
% So far, numerous recommendation systems have been proposed to learn user interests from implicit feedback. 
% Early methods primarily consisted of matrix factorization \cite{koren2009matrix} and factorization machines \cite{koren2008factorization}. And with the rapid advancement of deep learning, 
% models based on convolution neural networks(CNNs) \cite{tang2018personalized,yuan2019simple},  recurrent neural networks (RNNs) \cite{hidasi2015session,hidasi2016parallel}, and attention mechanisms \cite{kang2018self,sun2019bert4rec} have gained popularity in recommendation.

Recommender systems (RS) \cite{liu2018illuminating,cheng2021clue} have become crucial tools for information filtering in various real world scenarios, including e-commerce and media. They often use implicit feedback, such as clicks or purchases, due to its abundance, to infer user preferences \cite{wang2021denoising}. Initially, these systems used techniques like matrix factorization and factorization machines. The rise of deep learning has shifted the focus towards neural network-based models, including  convolution neural networks (CNNs) \cite{yuan2019simple}, recurrent neural networks (RNNs) \cite{hidasi2015session}, and those utilizing attention mechanisms \cite{kang2018self,sun2019bert4rec}, for more nuanced user interest profiling.
%many  deep neural networks based models have been widely proposed, including recurrent neural networks (RNNs) \cite{hidasi2015session,hidasi2016parallel}, convolution neural networks(CNNs) \cite{tang2018personalized,yuan2019simple},  and attention-based networks\cite{kang2018self,sun2019bert4rec}, have already achieved significant success and 
%become increasingly popular in recent years.

To train these deep recommendation systems, a common practice is to treat the recommendation task as a multi-class classification problem, in which the target item that the user interacted with often serves as the label.  Then, by combining the one-hot vector corresponding to the target item as supervisory signals, the models can be optimized through the application of Softmax loss. 
Unfortunately,  due to the randomness and diverse nature of the user interests, 
% and since the observed target items occupies only a minuscule portion of the entire item pool  \cite{pan2023understanding}, 
%	their behavior at one moment is often not limited to a single item. Consequently, 
the standard Softmax loss often fails to accurately reflect user interests and tends to ignore the difference between potential positive items and truly negative ones.
In recent years, some methods utilizing soft labels have been proposed to mitigate this issue. Label smoothing \cite{szegedy2016rethinking} introduces a uniform distribution to increase label smoothness, effectively preventing the recommender system from over-fitting. Furthermore, SoftRec \cite{cheng2021learning} suggests incorporating prior knowledge, such as item popularity distribution or user similarity, to generate more informative  soft targets for RS. Despite notable success, they directly optimize the joint distribution of target item and unobserved feedback items, 
which imposes inflexibility to model training due to the couple relation \cite{zhao2022decoupled} between such two kind items through our carefully theoretical analysis.
% which potentially imposes limitations and inflexibility on the optimization of both two kind items.
% modeling the latent interest distribution for unobserved feedback items. 

To address this challenge, in this paper, we introduce a novel decoupled soft label optimization framework for recommendation to separate  the soft objectives into two aspects, including \textbf{target confidence} and the \textbf{distribution of non-target items}. The former can be regarded as a binomial distribution related to the target, while the latter  is a $|\mathcal{I}|-1$ dimensional distribution reflecting the potential interest in unobserved feedback, where $\mathcal{I}$ is the item set. 
% Moreover, 
% through our theoretical analysis, we identify a couple relation between these two aspects in classical soft label optimization frameworks which directly use KL divergence and cross-entropy. As a result, it limits the flexible adjustment of target confidence and the weights of both components  \cite{zhao2022decoupled}, potentially leading model to sub-optimal performance. 
Based on which, a decoupled loss function is designed to flexibly adjust the weights of these two aspects. In addition, to maximize the performance of the proposed method, we also present a sensible soft label generation algorithm. In real life, user behaviors are often influenced by friends with similar preferences. Thus, we design an interest-sharing mechanism to explore users' potential positive feedback through their neighbors. We conduct extensive experiments on four recommendation models and three public datasets. The results demonstrate the effectiveness and generality of our proposed method.
We summarize the contributions as follows:
\begin{itemize}
	\item We propose a novel decoupled soft label optimization framework for recommendation.
 % to address that the widely used Softmax loss may lead to sub-optimal performance.
 It considers the objectives as two aspects by leveraging soft labels, including target confidence and the latent interest distribution of unobserved feedback. Based on which, a more flexibility loss function is designed to effectively adjust these two parts to achieve more accurate performance. 
	\item
	We present a sensible soft targets generation algorithm which leverages a label propagation process to capture user’s potential interest via neighbors.
	\item
	We conduct extensive experiments on various recommenders and datasets, the experimental results demonstrate the effectiveness  and generality of the proposed approach.
\end{itemize}
	\section{Related Work}
	\textbf{Learning Recommender System from Implicit Feedback.}
Typically, training a recommender system from implicit feedback \cite{wu2022multi,liu2018illuminating,han2023guesr} involves two main aspects: model architecture design and optimization objective formulation. This paper primarily  focuses on the latter.  Roughly,
existing loss functions can be categorized into three main types i.e. point wise binary cross entropy (BCE) loss \cite{pan2008one}, pair wise BPR loss \cite{rendle2012bpr} and Softmax \cite{sun2019bert4rec,blanc2018adaptive} loss which treats the problem as a multi-class classification task. As the Softmax loss has become one of the prevalent setting in recommendation, the utilized one-hot labels tends to overlook the distinctions in unobserved feedback and lead model to over-confidence. To this end, SoftRec \cite{cheng2021learning} replaces hard label with soft target to increase the label smoothness via prior distributions. Based on this, this paper delves into the study of the soft label learning framework and proposes a novel optimization perspective to enhance model learning from implicit feedback.
\\
\\
	\textbf{Evolving Recommender Models.}   
	Recommendation systems aim to predict positive feedback based on context. Early methods mainly involved matrix factorization \cite{koren2009matrix} and factorization machines \cite{koren2008factorization}. And with the rapid development of deep learning, neural networks based methods have been widely proposed for recommendation. YoutubeDNN \cite{covington2016deep} employs MLP to model diverse user features, while DeepFM \cite{guo2017deepfm} combines factorization machines  and deep neural networks  to learn both low-order and high-order feature combinations simultaneously. Besides, GRU4Rec \cite{hidasi2015session} proposes to model behavior sequence with recurrent neural networks (RNN) to capture the dynamic interests. And in recent years, self-attention models \cite{kang2018self} have shown their promising strengths in the capacity of long-term dependence modeling and easy parallel computation.

	\section{Method}
	% In this part, we first revisit the traditional soft label optimization framework in recommendation systems in section 4.1, and then provide a decouple perspective by dividing the objective into two aspects: the target confidence and the latent interest distribution of unobserved feedback. Based on which, a more flexible loss function and optimization framework is designed to effectively optimize these two aspects with soft targets. To maximize the effectiveness of the proposed approach, we also present a novel soft label generation algorithm which leverages a label propagation process to effectively capture the user's potential interest in unobserved items via neighbors in section 4.2.
\subsection{Problem Formulation}
	Assume that there are item set $\mathcal{I} = \{i_1, i_2, ..., i_{{m}}\}$ where $m$ is the number of items,
	%and user set $\mathcal{U} = \{u_1, u_2, ..., u_{|\mathcal{U}|}\}$, 
	the goal of recommender system is to learn a classifier $f_\theta(x):\mathcal{X} \rightarrow \mathcal{I}$. Where  $x$ is the input feature which can be represented by user or its historical interactions. Given the dataset $\mathcal{D} = \{(x_1,y_1),(x_2,y_2),...(x_{l},y_{l})\}$ where $y_j \in \mathcal{I}$ is the target item of $x_j$ and $l$ is the number of samples, denote the loss function as $\mathcal{L}$, the recommender training can be typically
	formalized as:
	\begin{equation}
		\theta^* =  \frac{1}{|\mathcal{D}|}argmin_\theta \sum_{(x,y) \in \mathcal{D}}\mathcal{L}(x,y),
	\end{equation}
	\subsection{Decoupled Soft Targets Optimization Framework}
% As previously discussed, the conventional soft target optimization framework typically seeks to optimize the joint distribution of the target item and unobserved feedback, which may introduce constraints and reduce flexibility in optimizing both types of items. To uncover the fundamental principles at play, next, we undertake a thorough examination of the classical framework and further propose a more adaptable loss function from a decouple perspective.
 
 Assume the model output after the Softmax operation is $p \in \mathbb{R}^{|\mathcal{I}|}$, denote the one-hot label and soft target as $d,q \in \mathbb{R}^{|\mathcal{I}|}$, respectively. Traditional optimization framework often trade off these two supervisory signals with a controlling parameter $\lambda_1 \in [0,1]$ as follow:
	\begin{equation}
		\mathcal{L}(p,q,d)=\lambda_1 \mathcal{D}_{KL}(q||p) + (1-\lambda_1) \mathcal{H}(d,p),
		\label{cekl}
	\end{equation}where $\mathcal{D}_{KL}$ and $\mathcal{H}$ stands for KL-divergence and cross-entropy. 
    \begin{figure}[t]
	\centering
	\includegraphics[width=0.9\linewidth]{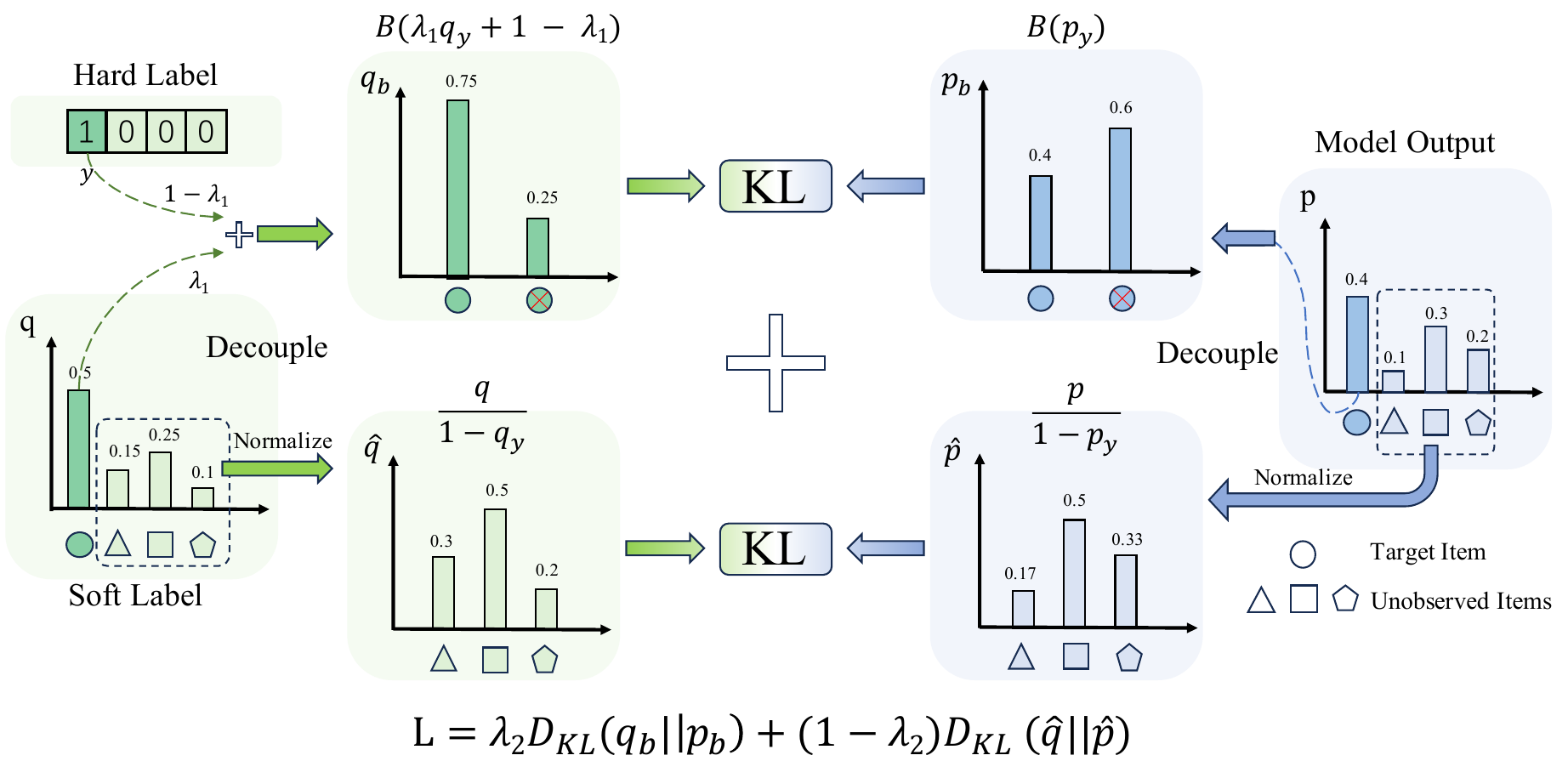}
	\caption{The proposed decoupled soft label optimization framework.}
	\label{model}
\end{figure}
 However, as previously discussed, the conventional soft target optimization framework typically optimize the joint distribution of the target item and unobserved feedback, which may introduce constraints and reduce flexibility in optimizing both types of items. To uncover the fundamental principles at play and provide a novel perspective, we separate the objective into these two aspects by the following theoretical analysis. Denote the target item as $y$, we have:
	\begin{equation}
		\begin{aligned}
			\mathcal{L}(p,q,d)
   % &=\lambda_1 \mathcal{D}_{KL}(q||p) + (1-\lambda_1) \mathcal{H}(d,p) \\
			%\\ &= \lambda_1  \sum_i q_ilogq_i -\lambda_1 \sum_i q_ilogp_i - (1-\lambda_1)logp_y \\
			&= \lambda_1 \sum_i q_ilogq_i -(\lambda_1 q_y +1 -\lambda_1) logp_y -\lambda_1  \sum_{i \neq y}  q_ilogp_i  
			\\&=\lambda_1 \sum_i q_ilogq_i -(\lambda_1 q_y +1 -\lambda_1) logp_y \\ &\;\;\;\;-\lambda_1 (1-q_y)\sum_{i \neq y} 
			\frac{q_i}{1-q_y} log \frac{p_i(1-p_y)}{1-p_y}
			\\ &=-(\lambda_1 q_y +1 -\lambda_1) logp_y -\lambda_1 \sum_{i \neq y}q_ilog(1-p_y)\\ &\;\;\;\;-\lambda_1 (1-q_y)\sum_{i \neq y} \frac{q_i}{1-q_y} log \frac{p_i}{1-p_y} + \lambda_1 \sum_i q_ilogq_i 
			\\&=-\left((\lambda_1 q_y +1 -\lambda_1) logp_y+ \lambda_1(1-q_y)log(1-p_y)\right) \\ &\;\;\;\;+\lambda_1 (1-q_y)\sum_{i \neq y}- \hat{q_i}log\hat{p_i} + \lambda_1 \sum_i q_ilogq_i 
			\\&= \mathcal{D}_{KL}(q_b||p_b) + \lambda_1(1-q_y) \mathcal{D}_{KL}(\hat{q}||\hat{p}) + F(q)%\lambda_1 \sum_{i \neq m} z_i log\frac{{z_i}}{\hat{z_i}} %\\&= KL(z_b||p_b) +\lambda_1(1-z_m)KL(\hat{z}||\hat{p}) + C(z)%-(1-z_m)log(1-z_m)
		\end{aligned}
            \label{classical}
	\end{equation}in which $q_b = Bernoulli(\lambda_1 q_y +1 -\lambda_1)$, $p_b = Bernoulli(p_y) \in \mathbb{R}^{2}$  can be regarded as the binomial distribution related to target item, while $\hat{p},\hat{q} \in \mathbb{R}^{|\mathcal{I}|-1}$  describes the distribution over unobserved feedback where $\hat{q}_i = q(i|i \neq y) = \frac{q_i}{q(i \neq y)} = \frac{q_i}{1-q_y} $, $\hat{p}_i =  p(i|i \neq y) = \frac{p_i}{p(i \neq y)} =  \frac{p_i}{1-p_y}\;(i\neq y) $, and $F(q)$ is a constant term unrelated to model. 
Actually, in Eq. \ref{classical}, $\mathcal{D}_{KL}(\hat{q}||\hat{p})$ reflecting the distribution over almost whole candidate set contains more information of the user's interests. However, according to the above analysis,
we can observe that the weight of the second part $\lambda_1(1-q_y)$ is directly constrained by $q_y$ while $q_y$ is often set to the highest value and suggested to be larger than $0.5$ \cite{cheng2021learning}. Moreover, it is inflexible to balance these two aspects since $\lambda_1$ is also used to control the confidence of the target item i.e. $Bernoulli(\lambda_1 q_y +1 -\lambda_1)$. Therefore, as shown in Figure \ref{model}, we propose to decouple the objectives by re-weighting the loss function as:
	\begin{equation}
		\begin{aligned}
			\mathcal{L}_{Deso}(p,q,d) &=  \lambda_2 \mathcal{D}_{KL}(q_b||p_b) + (1-\lambda_2) \mathcal{D}_{KL}(\hat{q}||\hat{p}),
			\label{lnew}
		\end{aligned}
	\end{equation}
where $\lambda_2$ is a trade-off parameter. As a result, it is allowed to effectively adjust the
	importance of these two aspects while accurately modeling the binomial distribution related to the target confidence.
	\subsection{Soft Target Generation}
	% One-hot labels tends to ignore the difference between potential
	% positive items and truly negative ones, leading model to over-confidence.  
	% Although some methods address this challenge through soft labels, most of them simply introduce some prior distributions, such as uniformity, item popularity, or user similarity, still fails short in capturing the potential user preference among unobserved feedback. To maximize the effectiveness of the proposed approach, we further conduct a label propagation process to explore potential positive feedback via neighbors in this part.
	\subsubsection{Neighbor-based Graph}
	In real life, user's behaviors are often tend to be inspired by others who share similar interests. Hence, utilizing neighbor information, like social networks, aids in mitigating data sparsity and capturing more accurate user preferences. However, since many recommendation systems lack such type of data, we replace it with employing the K-means algorithm to cluster pretrained embedding to acquire the neighbors for each user.
	Then, any two points within the same cluster can be regarded as neighbors i.e. 
		$N(u) = \{v | C(u)=C(v)\},$
	where $N(u)$ is the neighbor set of $u$ and $C(v)$ represents the cluster to which $v$ belongs. Besides, the similarity of users can be simply measured by the l2-distance as:
		$sim(u,v) = {-||E_u - E_v||_2},$
	\subsubsection{Label Propagation}
	We initialize each user $u$  with a one-hot label $q_u^0 = d^u$. And in each label propagation round, all of them will receive distributions  from their neighbors to update their current soft targets. As similar users are more likely to share their information, we employ normalized similarity with a temperature parameter $\tau$ to measure the propagation weight as:
 	\begin{equation}
		w_{uv} = p(v|u) = \frac{exp({sim(u,v)}/\tau)}{\sum_{j \in N(u)}exp(sim(u,j)/\tau)}
  \label{simw}
	\end{equation}
	After that, the distribution update process for each user can be represented as:
	\begin{equation}
		q_u^{t+1} =  \sum_{v \in N(u)}p(v|u) q^t_v = w_uq^t
	\end{equation}
	To retain the original information and ensure that the probability of the target is maximized which suggested in \cite{cheng2021learning}, we also keep the user's raw target with a probability of $0.5$. Thus, the overall label update process can be rewritten as:
	\begin{equation}
		q^{t+1} =  \frac{1}{2} wq^t + \frac{1}{2}q^0.
	\end{equation}
	We can get the final soft targets $q^*=q^T$ after $T$ iterations.	
	Overall, we present our decoupled soft target
	optimization framework DeSoRec in Algorithm \ref{al}.
		\begin{algorithm}[t]
		\caption{Decoupled soft target optimization framework} 
		\label{al} 
		\begin{algorithmic}[1]
			%		\item[]
			\REQUIRE  recommender system $f_\theta$, dataset $\mathcal{D}=(x_i,y_i)_{i=1}^n$, number of clusters $k$, temperature parameter $\tau$.
			\ENSURE soft targets $q^*$, well-trained model $\theta^*$.
			\STATE Initial RS parameter $\theta$
			\STATE $\theta_0 = arg\;min_\theta\ \sum_{i=1}^n\mathcal{H}(y_i,f_\theta(x_i))$ // Pretrain the RS 
			\STATE $\{c_1,c_2,...c_k\} = K\_means(f_{\theta_0}(x),k)$
			\FOR{$c = c_1, c_2, ... c_k$}
			\STATE Compute the similarity matrix $w_c(u,v)$ of users in $c$  according to Eq. \ref{simw}
			\STATE $q^0_c = d_c = Onehot(Y_c)$ // Initial the label distribution with target item
			\FOR{$t = 0,1,2,...T-1$}
			\STATE 	$	q^{t+1}_c  \gets  \frac{1}{2} wq^t_c + \frac{1}{2}q^0_c $
			\ENDFOR
			\STATE Get the final soft targets of $c$ as $q^*_c = q^T_c$
			\ENDFOR 
			\STATE $q^* = [q^*_1,q^*_2,...q^*_k],$\;\;$\theta^* = arg\;min_\theta\ \sum_{i=1}^n\mathcal{L}_{Deso}(f_\theta(x_i),q^*,d_i)$ 
		\end{algorithmic}  
	\end{algorithm}
	\section{Experiments}
 % \vspace{-0.12in}
	\subsection{Experimental Settings}
 	% \vspace{-0.08in}
	\subsubsection{Datasets}
	We conduct experiments on three dataset from various domains including transaction \textbf{(Diginetica\footnote{\url{http://cikm2016.cs.iupui.edu/cikm-cup}})}, movie \textbf{(MovieLens\footnote{\url{http://files.grouplens.org/datasets/movielens/}})} and music \textbf{(Last.FM\footnote{\url{www.dtic.upf.edu/ocelma/MusicRecommendationDataset/lastfm-1K.html}})}.
% \textbf{Diginetica\footnote{\url{http://cikm2016.cs.iupui.edu/cikm-cup}}}: is a challenge dataset for CIKM cup 2016 which
% 		contains the transaction data. And we take the user click logs into account.
% \textbf{MovieLens\footnote{\url{http://files.grouplens.org/datasets/movielens/}}}: is a widely used benchmark movie dataset for evaluating recommender systems.
% \textbf{Last.FM\footnote{\url{www.dtic.upf.edu/ocelma/MusicRecommendationDataset/lastfm-1K.html}}}. is a well-known music recommendation dataset. It contains information about the songs and artists that users have listened to. %And sessions where the time span between the last two items is longer than 2 hours are ignored.
		% \item[] \textbf{Wiki}: this is a  recommendation dataset that contains user historical interactions with Wikipedia articles.
		% Table generated by Excel2LaTeX from sheet 'Sheet1'
	For all datasets, users and items that interacted less than five times are filtered out. And the max length of user behaviors is set to 20. The statics of these datasets after processing are summarized in Table \ref{data}.
	% \vspace{-0.2in}
	\subsubsection{Testing Recommenders and Compared Methods}
	We conduct our experiments on four modern recommenders, including two feature based models \textbf{DeepFM} \cite{guo2017deepfm} and \textbf{YoutubeDNN} \cite{covington2016deep}, two sequential recommenders \textbf{GRU4Rec} \cite{hidasi2015session} and \textbf{SASRec} \cite{kang2018self}. We compare our method with standard softmax loss (\textbf{Base}) and label smoothing (\textbf{LS}) \cite{szegedy2016rethinking}. Besides, we also choose two of the most competitive methods, \textbf{POP+} and \textbf{CSN} from \cite{cheng2021learning} as baselines, which design soft labels based on item popularity strategy and user strategy, respectively.
	% \vspace{-0.2in}
	\subsubsection{Implementation Details}
	To ensure fairness in our experiments, we keep the consistent model parameters for all methods. We set the embedding size to 128 and conduct the Adam optimizer with a learning rate of $1e^{-3}$, while maintaining a batch size of 256. We carefully tune the trade-off parameter $\lambda_1$ of soft target and hard label in baselines. As for our proposed method, we perform a grid search to find the best temperature and the number of class from $\{0.25,0.5,1.0,2.0\}$ and $\{16,64,256,1024\}$.  Note that we tune trade-off weight $\lambda_1$ and $\lambda_2$ from $\{0.1,0.3,0.5,0.7,0.9\}$.
  	 	\begin{table}[t]
	\centering
	\caption{Statics of the used datasets in the experiments.}   
	\tabcolsep=0.1cm
	\renewcommand{\arraystretch}{0.9}
	\resizebox{1\columnwidth}{!}{
		\begin{tabular}{cccccc}
			\toprule
			Datasets & \multicolumn{1}{l}{\#Num. Users} & \multicolumn{1}{l}{\#Num. Items} & \multicolumn{1}{l}{\#Num. Actions} & \multicolumn{1}{l}{\#Actions/User} & \multicolumn{1}{l}{\#Actions/Item} \\
			\midrule
			Diginetica & 62,502 & 43,229 & 467,631 & 7.48  & 10.81  \\
			Movielens & 1,460,900 & 23,716 & 27,525,126 & 18.84  & 1160.61  \\
			Last.FM & 534,982 & 199,013 & 10,699,640 & 20.00  & 53.76  \\
			% Wiki & 1,070 & 4,911 & 18,778 & 17.55  & 3.82  \\
			\bottomrule
		\end{tabular}%
	}
	\label{data}%
\end{table}%
	\subsubsection{Evaluation Metrics}
	We evaluate the recommendation performance over all candidate items with \textit{Recall} and  \textit{Normalized Discounted
		Cumulative Gain (NDCG)}. 
	%and \textit{Mean Reciprocal Ranking (MRR)}.
	The first one is an evaluation of unranked retrieval sets while the other reflects the order of ranked lists. We consider top-k of overall item set for recommendations where $k \in \{20,10\}$ in experiments. And we split the dataset into training, validation and testing sets following the leave-one-out strategy described in \cite{zhao2020revisiting}.
	% Table generated by Excel2LaTeX from sheet 'main'
		\begin{table*}[t]
		\centering
		\caption{Recommendation performance comparison of different methods over four recommenders and three datasets. The best performance and the second best performance methods are indicated by bold and underlined fonts.}
		\tabcolsep=0.1cm  
		\renewcommand{\arraystretch}{1.15}
		\resizebox{1\textwidth}{!}{
			\begin{tabular}{cccccccccccccc}
				\toprule
				\multirow{2}[2]{*}{Recommenders} & \multirow{2}[2]{*}{Methods} & \multicolumn{4}{c}{Diginetica} & \multicolumn{4}{c}{MovieLens} & \multicolumn{4}{c}{Last.FM} \\
				&   & R@20 & N@20 & R@10 & N@10 & R@20 & N@20 & R@10 & N@10 & R@20 & N@20 & R@10 & N@10 \\
				\midrule
				\multirow{4}[2]{*}{DeepFM} & Base & 0.1507  & \underline{0.0761}  & 0.1124  & \underline{0.0665}  & 0.2390  & 0.1099  & 0.1652  & 0.0913  & 0.2738  & 0.2024  & 0.2463  & 0.1955  \\
				& LS & \underline{0.1508}  & 0.0760  & \underline{0.1125}  & 0.0663  & 0.2418  & 0.1118  & 0.1679  & 0.0933  & 0.2931  & \underline{0.2142}  & \underline{0.2632}  & \textbf{0.2066} \\
				& POP+ & 0.1418  & 0.0724  & 0.1071  & 0.0636  & \underline{0.2450}  & \underline{0.1133}  & \underline{0.1705}  & \underline{0.0945}  & \underline{0.2938}  & 0.2134  & 0.2629  & 0.2056  \\
    				& CSN & 0.1500  & 0.0753  & 0.1111  & 0.0655  & 0.2443  & 0.1127  & 0.1691  & 0.0937  & 0.2841  & 0.2041  &  0.2533 & 0.1964   \\
				& Ours & \textbf{0.1581} & \textbf{0.0792} & \textbf{0.1163} & \textbf{0.0686} & \textbf{0.2546} & \textbf{0.1189} & \textbf{0.1778} & \textbf{0.0995} & \textbf{0.3035} & \textbf{0.2150} & \textbf{0.2689} & \underline{0.2063}  \\
				\midrule
				\multirow{4}[2]{*}{YoutubeDNN} & Base & 0.1350  & 0.0753  & 0.1077  & 0.0684  & 0.2427  & 0.1125  & 0.1689  & 0.0939  & 0.2028  & 0.1590  & 0.1850  & 0.1545  \\
				& LS & 0.1402  & 0.0819  & 0.1151  & 0.0755  & 0.2473  & 0.1149  & 0.1720  & 0.0961  & 0.2409  & 0.1950  & 0.2232  & 0.1905  \\
				& POP+ & 0.1334  & 0.0789  & 0.1107  & 0.0732  & 0.2482  & 0.1155  & 0.1729  & 0.0965  & 0.2428  & \underline{0.1962}  & 0.2242  & \underline{0.1913}  \\
    				& CSN & \underline{0.1521}  & \underline{0.0846}  & \underline{0.1213}  & \underline{0.0768}  & \underline{0.2519}  & \underline{0.1177}  & \underline{0.1763}  & \underline{0.0987}  & \underline{0.2465}  & 0.1947  & \underline{0.2255}  & 0.1894  \\
				& Ours & \textbf{0.1597} & \textbf{0.0870} & \textbf{0.1247} & \textbf{0.0782} & \textbf{0.2578} & \textbf{0.1204} & \textbf{0.1803} & \textbf{0.1009} & \textbf{0.2669} & \textbf{0.2085} & \textbf{0.2442} & \textbf{0.2024} \\
				\midrule
				\multirow{4}[2]{*}{GRU4Rec} & Base & 0.1066  & 0.0505  & 0.0767  & 0.0430  & 0.2473  & 0.1136  & 0.1705  & 0.0942  & 0.2558  & 0.2024  & 0.2345  & 0.1970  \\
				& LS & 0.1173  & 0.0557  & 0.0843  & 0.0473  & 0.2488  & 0.1153  & 0.1725  & 0.0942  & 0.2725  & 0.2152  & 0.2499  & 0.2094  \\
				& POP+ & \underline{0.1197}  & \underline{0.0591}  & \underline{0.0881}  & \underline{0.0511}  & 0.2471  & 0.1143  & 0.1714  & \underline{0.0952}  & \underline{0.2792}  & \underline{0.2177}  & \underline{0.2552}  & \underline{0.2117}  \\
    				& CSN & 0.1181  & 0.0570  & 0.0853  & 0.0487  & \underline{0.2517}  & \underline{0.1160}  & \underline{0.1740} & 0.0945  & 0.2668  & 0.2102  & 0.2442  & 0.2045  \\
				& Ours & \textbf{0.1289} & \textbf{0.0623} & \textbf{0.0942} & \textbf{0.0536} & \textbf{0.2586} & \textbf{0.1182} & \textbf{0.1777} & \textbf{0.0978} & \textbf{0.2834} & \textbf{0.2199} & \textbf{0.2578} & \textbf{0.2134} \\
				\midrule
				\multirow{4}[2]{*}{SASRec} & Base & 0.1766 & 0.1075 & 0.1517 & \underline{0.1012} & 0.2612  & 0.1298  & 0.1905  & 0.1120  & 0.3248  & 0.2608  & 0.3084  & 0.2567  \\
				& LS & 0.1818 & \underline{0.1088} & 0.1538 & \textbf{0.1017} & 0.2696  & \underline{0.1343}  & 0.1969  & \underline{0.1159}  & \underline{0.3287}  & 0.2626  & \underline{0.3114}  & 0.2582  \\
				& POP+ & 0.1750 & 0.1067 & 0.1501 & 0.1004 & \underline{0.2700}  & 0.1340  & \underline{0.1972}  & 0.1157  & 0.3156  & 0.2676  & 0.3016  & 0.2641  \\
    				& CSN &\underline{0.1843} & 0.1073 & \underline{0.1550} & 0.0999 & 0.2651  & 0.1319  & 0.1938  & 0.1140  & 0.3241  & \underline{0.2684}  & 0.3082  & \underline{0.2643}  \\
				& Ours & \textbf{0.1895} & \textbf{0.1093} & \textbf{0.1562} & 0.1009 & \textbf{0.2733} & \textbf{0.1351} & \textbf{0.1983} & \textbf{0.1162} & \textbf{0.3360} & \textbf{0.2735} & \textbf{0.3157} & \textbf{0.2684} \\
				\bottomrule
		\end{tabular}}%
		\label{t1}%
  \vspace{-0.2in}
	\end{table*}%
 \vspace{-0.3in}
 		\subsection{Performance Comparison}
   \vspace{-0.1in}
   The performance results of different methods on various benchmark datasets and recommenders are shown in Table \ref{t1}. 
   % We observe that LS achieved better performance compared to traditional optimization objectives that only use one-hot labels, confirming the necessity of learning with soft targets. POP+ and CSN obtains further improvements by designing soft targets that more suitable for recommendation. 
   Firstly, we observe that LS, POP+ and CSN achieves better performance compared to Base, confirming the necessity of learning with soft targets. 
% However, these method directly optimize 
% % the model with KL and cross-entropy loss function,
% the joint distribution of the target item and unobserved feedback,
% potentially bringing inflexibility in both type of items through the theoretical analysis above. 
Moreover, different from these methods, we introduce a novel decoupled soft target optimization framework for recommendation, which allows to flexibly adjust the weights of the target confidence and non-target items. As a result, DeSoRec performs best on most benchmark datasets and models, demonstrating the effectiveness and generality of the proposed method.
\vspace{-0.3in}
\subsection{Generality Analysis with Other Soft Label}
\vspace{-0.1in}
We investigate the wide-ranging utility of our decoupled optimization framework by integrating it with alternative soft label generation techniques. As documented in Table \ref{deso}, the results clearly demonstrate that our optimization framework is universally effective across different soft label generation methods. Moreover, we find that our framework further enhancements to the label propagation algorithm, underscoring the capacity of our soft label generation approach.

\begin{table}[t]
	\centering
	\caption{Recommendation performance (NDCG@10) of other soft targets methods with decoupled learning on DeepFM. $\Delta$ means the improvement.}
	\tabcolsep=0.5cm  
			\resizebox{1\linewidth}{!}{
		\begin{tabular}{c|ccc|ccc}
			\toprule
   & \multicolumn{3}{c}{Diginetica} & \multicolumn{3}{c}{MovieLens} \\
			Method & w/o De & w/ De & $\Delta$ & w/o De & w/ De & $\Delta$ \\
			\midrule
			LS& 0.0663 & \textbf{0.0680} & 2.56\% & 0.0933 & \textbf{0.0952} & 2.04\%  \\
                POP+& 0.0636 & \textbf{0.0655} & 2.99\% & 0.0945 & \textbf{0.0959} & 1.48\% \\
                CSN& 0.0655 & \textbf{0.0671} & 2.44\% & 0.0937 & \textbf{0.0957} & 2.13\% \\
                DeSoRec& 0.0666 & \textbf{0.0686} & 3.00\% & 0.0964 & \textbf{0.0995} & 3.22\%\\
			% \multirow{6}[6]{*}{MovieLens} & LS & 0.2418 & 0.1118 & 0.1679 & 0.0933 \\
			% & LS + De & 0.2440 & 0.1138 & 0.1704 & 0.0952 \\
			% & $\Delta$ & 0.91\% & 1.79\% & 1.49\% & 2.04\% \\ \cline{2-6}  \rule{0pt}{3ex}
			% &  SoftRec & 0.2444 & 0.1135 & 0.1704 & 0.0949 \\
			% & SoftRec + De & 0.2444 & 0.1140 & 0.1711 & 0.0959 \\
			% & $\Delta$ & 0.00\% & 0.40\% & 0.41\% & 1.05\% \\ \cline{2-6}  \rule{0pt}{3ex}
			% &  DeSoRec - De & 0.2497 & 0.1157 & 0.1733 & 0.0964 \\
			% & DeSoRec & 0.2546 & 0.1189 & 0.1778 & 0.0995 \\
			% & $\Delta$ & 1.96\% & 2.77\% & 2.60\% & 3.22\%  \\
			\bottomrule
		\end{tabular}%
  }
		\label{deso}%
	\end{table}%
 \begin{table}[t]
		\centering
		\caption{Ablation study of each decoupled objectives on YoutubeDNN.}
		\tabcolsep=0.5cm  
    \renewcommand{\arraystretch}{1}
    	\resizebox{1\linewidth}{!}{
    \begin{tabular}{lcccc}
    \toprule
    \multirow{2}[2]{*}{Method} & \multicolumn{2}{c}{Diginetica} & \multicolumn{2}{c}{Last.FM} \\
      & {R@10} & {N@10} & {R@10} & {N@10} \\
    \midrule
    Base & 0.1077 & 0.0684 & 0.1850 & 0.1545 \\
    Target Confidence only  & 0.1145 & 0.0716 & 0.1932 & 0.1603 \\
    Non-target Distribution only & 0.0753 & 0.0432 & 0.1138 & 0.0736 \\
    DeSoRec & \textbf{0.1247} & \textbf{0.0782} & \textbf{0.2442} & \textbf{0.2024} \\
    \bottomrule
    \end{tabular}}%
  \label{remove}%
  	\vspace{-0.2in}
\end{table}%

 \vspace{-0.15in}
	\subsection{Abaltion Study}
	\subsubsection{Effectiveness of  Each Decoupled Objectives.}
	To examine the effectiveness of the two decoupled optimization objectives independently, 
	we conducted experiments focusing separately on the target confidence and the non-target distributions.
	 As indicated in Table \ref{remove}, optimizing only the target confidence outperforms Base, likely because replacing one-hot labels with a soft binomial distribution helps prevent the model from becoming over-confident. Conversely, optimization focus only on the non-target items delivers the least favorable outcomes, highlighting the critical importance of capturing user interest through positive feedback. Notably, we can observe that removing either part results in a performance decrease, emphasizing the necessity of jointly addressing both aspects and capturing user's potential interest in items without observed feedback.
	% Table generated by Excel2LaTeX from sheet 'ablation'
	% \begin{table}[t]
	% 	\centering
	% 	\caption{Ablation study of the effectiveness of each part of the loss function. Experiments are conducted  using YoutubeDNN.}
	% 	\tabcolsep=0.2cm  
 %    \renewcommand{\arraystretch}{0.9}
 %    	\resizebox{1\linewidth}{!}{
	% 		\begin{tabular}{clcccc}
	% 			\toprule
	% 			Datasets & \multicolumn{1}{c}{Method} & R@20 & N@20 & R@10 & N@10 \\
	% 			\midrule
	% 			\multirow{4}[2]{*}{Diginetica} & Base & 0.1350  & 0.0753  & 0.1077  & 0.0684  \\
	% 			& Target Confidence only ($\lambda_2=1$) & 0.1440  & 0.0790  & 0.1145  & 0.0716  \\
	% 			& Non-target Distribution only ($\lambda_2=0$) & 0.1007  & 0.0496  & 0.0753  & 0.0432  \\
	% 			& DeSoRec & \textbf{0.1597} & \textbf{0.0870} & \textbf{0.1247} & \textbf{0.0782} \\
	% 			\midrule
	% 			\multirow{4}[2]{*}{Last.FM} & Base & 0.2028  & 0.1590  & 0.1850  & 0.1545  \\
	% 			& Target Confidence only ($\lambda_2=1$) & 0.2103  & 0.1648  & 0.1932  & 0.1603  \\
	% 			& Non-target Distribution only  ($\lambda_2=0$) & 0.1364  & 0.0793  & 0.1138  & 0.0736  \\
	% 			& DeSoRec & \textbf{0.2669} & \textbf{0.2085} & \textbf{0.2442} & \textbf{0.2024} \\
	% 			\bottomrule
	% 		\end{tabular}}%
	% 	\label{remove}%
	% \end{table}%

 % Table generated by Excel2LaTeX from sheet 'Sheet1'

\vspace{-0.15in}
	\subsubsection{In-depth Analysis of DeSoRec's Flexibility.}
\begin{figure}[t]
	\centering
	\includegraphics[width=1\linewidth]{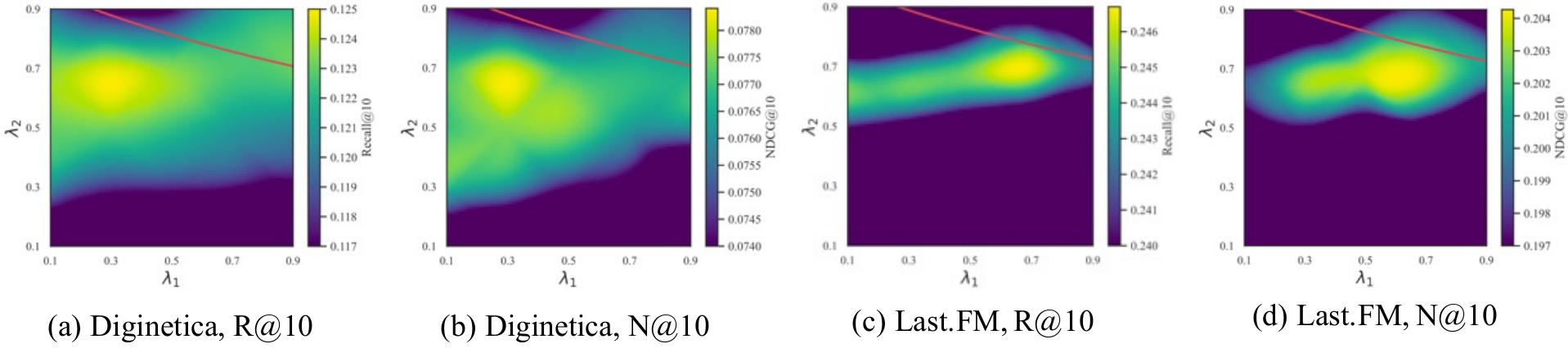}
 \vspace{-0.3in}
	\caption{Analysis of the trade-off parameters. Experiments are conducted  using YoutubeDNN. We search $\lambda_1$, $\lambda_2$ from \{0.1,0.3,0.5,0.7,0.9\} and adopt interpolation to fill the parameter space. The red line $\lambda_2 = \frac{1}{1+\lambda_1(1-\overline{q_y})}$ represents the search space of the traditional optimization framework coupling $\lambda_1$ and $\lambda_2$.}
	\label{ld12}
 \vspace{-0.2in}
\end{figure}

	As mentioned earlier, classical optimization frameworks using soft labels face challenges in adjusting the weights of two objectives flexibly. 
 % This is because that the weight parameter $\lambda_1$ is  coupled  with the binomial distribution for target confidence $Bernoulli(\lambda_1 q_y +1 -\lambda_1)$. 
 To thoroughly explore the combined impact of \( \lambda_1 \) and \( \lambda_2 \) on recommendation performance, we independently adjust them from \( \{0.1, 0.3, 0.5, 0.7, 0.9\} \), and then fill the remaining parameter space using interpolation. The results under YoutubeDNN are illustrated in Figure \ref{ld12}. From this, we can observe that both excessively large and small settings lead to performance degradation. The optimal \( \lambda_1 \) for Diginetica and LastFM is around \( 0.3 \) and \( 0.6 \), respectively, while the optimal \( \lambda_2 \) falls between \( 0.5 \) and \( 0.7 \) for both dataset. However, the parameter search range of traditional soft label optimization framework is approximately equivalent to being restricted on the red line i.e.  \( \lambda_2 = \frac{1}{1+\lambda_1(1 - 
\overline{q_y})} \) by comparing Eq. \ref{classical} and Eq. \ref{lnew}, where $\overline{q_y}$ is the average target confidence. As a result, it limits them to sub-optimal results as the best parameter groups lie in the blind zone. This demonstrates that the decoupled optimization of target and remaining items can effectively balance the weights of these two aspects while accurately modeling the target confidence.
% \begin{figure}[t]
% 		\centering
% %		\hspace{-0.1cm}
% 		%		\includegraphics[width=1\columnwidth]{image/sensitivity_recall.pdf}
% 			\hspace{-2.5mm}
% 		\subfigure[\scriptsize{Diginetica, R@10}]{
% 			\centering
% 			\includegraphics[width=0.26\linewidth]{image/1deg_ld1ld2_R.pdf}
% 		} 
% 	\hspace{-6mm}
% 		\subfigure[\scriptsize{Diginetica, N@10}]{
% 			\centering
% 			\includegraphics[width=0.26\linewidth]{image/1deg_ld1ld2_N.pdf}
% 		}
% 	\hspace{-6mm}
% 		\subfigure[\scriptsize{Last.FM, R@10}]{
% 			\centering
% 			\includegraphics[width=0.26\linewidth]{image/1lastfm_ld1ld2_R.pdf}
% 		}
% 		\hspace{-6mm}
% 		\subfigure[\scriptsize{Last.FM, N@10}]{
% 			\centering
% 			\includegraphics[width=0.26\linewidth]{image/1lastfm_ld1ld2_N.pdf}
% 		}
% 		\caption{Analysis of the trade-off parameters. Experiments are conducted  using YoutubeDNN. We search $\lambda_1$, $\lambda_2$ from \{0.1,0.3,0.5,0.7,0.9\} and adopt interpolation to fill the parameter space. The red line $\lambda_2 = \frac{1}{1+\lambda_1(1-\overline{q_y})}$ represents the search space of the traditional optimization framework coupling $\lambda_1$ and $\lambda_2$.  }
%   \vspace{-0.25in}
% 		\label{ld12}
% 	\end{figure}

\vspace{-0.2in}
	\section{Conclusion}
 \vspace{-0.1in}
In this paper, we introduced a novel decoupled soft target optimization framework for recommendation. Specifically, we treated the optimization of recommender systems as  two parts: the target confidence and the distributions of unobserved feedback, and further theoretically design a new loss function to flexibly adjust the importance of two aspects. To maximize the effectiveness of the proposed method, we also presented a sensible soft label generation algorithm which utilizes a label propagation process to capture users' potential interests via neighbors. Finally, we conducted expensive experiments on various public datasets and four popular recommender models. The experimental results showed that the proposed approach could obtain a higher performance gain than competitive baselines.
 We hope this paper could inspire more works to be proposed from the optimization perspective for recommender system.\\\\
 \textbf{Acknowledgement} This research was supported by grants from the National Key Research and Development Program of China (Grant No. 2021YFF0901003) and the Fundamental Research Funds for the Central Universities.
% \*vspace{-0.2in}
	\begin{small}
		\bibliographystyle{splncs04}
		\bibliography{ref}
	\end{small}
\end{document}